\documentclass{elsart}

\usepackage{graphicx}
\usepackage{epsfig}
\usepackage{amssymb}
\usepackage{url}
\begin{document}

\begin{frontmatter}
\title{ARGO-YBJ constraints on very high energy emission from GRBs}
\author{\centerline {{\it The ARGO-YBJ Collaboration:}}}
\newline
\author[ROMA2,INFN_RM2]{G. Aielli},
\author[ROMA3,INFN_RM3]{C. Bacci},
\author[INFN_NA,NAPOLI]{B. Bartoli},
\author[LECCE,INFN_LE]{P. Bernardini},
\author[IHEP] {X.J. Bi},
\author[LECCE,INFN_LE]{C. Bleve},
\author[INFN_RM3]{P. Branchini},
\author[INFN_RM3]{A. Budano},
\author[ROMA3,INFN_RM3]{S. Bussino},
\author[INFN_VBP]{A.K. Calabrese Melcarne},
\author[ROMA2,INFN_RM2]{P. Camarri},
\author[IHEP]{Z. Cao},
\author[INFN_TO,IFSI]{A. Cappa},
\author[INFN_RM2]{R. Cardarelli},
\author[NAPOLI,INFN_NA]{S. Catalanotti},
\author[INFN_PV]{C. Cattaneo},
\author[ROMA3,INFN_RM3]{P. Celio},
\author[IHEP]{S.Z. Chen \corauthref{cor1}},
\corauth[cor1]{Corresponding author. Tel: +86 10 88236106; Fax: +86
10 88233086} \ead{chensz@ihep.ac.cn}
\author[IHEP]{Y. Chen},
\author[IHEP]{N. Cheng},
\author[INFN_LE]{P. Creti},
\author[HEBEI] {S.W. Cui},
\author[YUNNAN]{B.Z. Dai},
\author[INFN_CT,DIFTER]{G. D'Al\'{\i} Staiti},
\author[TIBET]{Danzengluobu},
\author[INFN_TO,IFSI,FISTO]{M. Dattoli},
\author[LECCE,INFN_LE]{I. De Mitri},
\author[NAPOLI,INFN_NA]{B. D'Ettorre Piazzoli},
\author[ROMA3,INFN_RM3]{M. De Vincenzi},
\author[NAPOLI,INFN_NA]{T. Di Girolamo},
\author[TIBET]{X.H. Ding},
\author[INFN_RM2]{G. Di Sciascio},
\author[SHANDONG]{C.F. Feng},
\author[IHEP]{Zhaoyang Feng},
\author[CHENGDU]{Zhenyong Feng},
\author[INFN_RM3]{F. Galeazzi},
\author[INFN_TO,FISTO]{P. Galeotti},
\author[INFN_RM3]{R. Gargana},
\author[IHEP]{Q.B. Gou},
\author[IHEP]{Y.Q. Guo},
\author[IHEP]{H.H. He},
\author[TIBET]{Haibing Hu},
\author[IHEP]{Hongbo Hu},
\author[CHENGDU]{Q. Huang},
\author[NAPOLI,INFN_NA]{M. Iacovacci},
\author[ROMA2,INFN_RM2]{R. Iuppa},
\author[ROMA3,INFN_RM3]{I. James},
\author[CHENGDU]{H.Y. Jia},
\author[TIBET]{Labaciren},
\author[TIBET]{H.J. Li},
\author[SHANDONG]{J.Y. Li},
\author[IHEP]{X.X. Li},
\author[INFN_RM2]{B. Liberti},
\author[INFN_PV,PAVIA]{G. Liguori},
\author[IHEP]{C. Liu},
\author[YUNNAN]{C.Q. Liu},
\author[HEBEI]{M.Y. Liu},
\author[YUNNAN]{J. Liu},
\author[IHEP]{H. Lu},
\author[IHEP]{X.H. Ma},
\author[LECCE,INFN_LE]{G. Mancarella},
\author[ROMA3,INFN_RM3]{S.M. Mari},
\author[INFN_LE,ING_INNO]{G. Marsella},
\author[LECCE,INFN_LE]{D. Martello},
\author[INFN_NA]{S. Mastroianni},
\author[TIBET]{X.R. Meng},
\author[ROMA3,INFN_RM3]{P. Montini},
\author[TIBET]{C.C. Ning},
\author[INFN_CT,IASF_BO]{A. Pagliaro},
\author[INFN_LE,ING_INNO]{M. Panareo},
\author[INFN_LE,ING_INNO]{L. Perrone},
\author[ROMA3,INFN_RM3]{P. Pistilli},
\author[SHANDONG] {X.B. Qu},
\author[INFN_NA]{E. Rossi},
\author[INFN_RM3]{F. Ruggieri},
\author[NAPOLI,INFN_NA]{L. Saggese},
\author[INFN_PV]{P. Salvini},
\author[ROMA2,INFN_RM2]{R. Santonico},
\author[IHEP]{P.R. Shen},
\author[IHEP]{X.D. Sheng},
\author[IHEP]{F. Shi},
\author[INFN_RM3]{C. Stanescu},
\author[INFN_LE]{A. Surdo},
\author[IHEP]{Y.H. Tan},
\author[INFN_TO,IFSI]{P. Vallania},
\author[INFN_TO,IFSI]{S. Vernetto},
\author[INFN_TO,FISTO]{C. Vigorito},
\author[IHEP]{B. Wang},
\author[IHEP]{H. Wang},
\author[IHEP]{C.Y. Wu},
\author[IHEP]{H.R. Wu},
\author[CHENGDU] {B. Xu},
\author[SHANDONG]{L. Xue},
\author[HEBEI]{Y.X. Yan},
\author[YUNNAN]{Q.Y. Yang},
\author[YUNNAN]{X.C. Yang},
\author[TIBET]{A.F. Yuan},
\author[IHEP]{M. Zha},
\author[IHEP]{H.M. Zhang},
\author[IHEP]{JiLong Zhang},
\author[IHEP]{JianLi Zhang},
\author[YUNNAN]{L. Zhang},
\author[YUNNAN]{P. Zhang},
\author[SHANDONG]{X.Y. Zhang},
\author[IHEP]{Y. Zhang},
\author[TIBET]{Zhaxisangzhu},
\author[CHENGDU]{X.X. Zhou},
\author[IHEP]{F.R. Zhu},
\author[IHEP]{Q.Q. Zhu},
\author[LECCE,INFN_LE]{G. Zizzi}

\address[ROMA2]{Dipartimento di Fisica dell'Universit\`a ``Tor Vergata'',
                 via della Ricerca Scientifica 1, 00133 Roma, Italy}
\address[INFN_RM2]{Istituto Nazionale di Fisica Nucleare, Sezione di Tor Vergata,
         via della Ricerca Scientifica 1, 00133 Roma, Italy}
\address[ROMA3]{Dipartimento di Fisica dell'Universit\`a ``Roma Tre'',
                 via della Vasca Navale 84, 00146 Roma, Italy}
\address[INFN_RM3]{Istituto Nazionale di Fisica Nucleare, Sezione di
                   Roma3, via della Vasca Navale 84, 00146 Roma, Italy}
\address[INFN_NA]{Istituto Nazionale di Fisica Nucleare, Sezione di
                 Napoli, Complesso Universitario di Monte
                 Sant'Angelo, via Cinthia, 80126 Napoli, Italy}
\address[NAPOLI]{Dipartimento di Fisica dell'Universit\`a di Napoli,
                 Complesso Universitario di Monte
                 Sant'Angelo, via Cinthia, 80126 Napoli, Italy}
\address[LECCE]{Dipartimento di Fisica dell'Universit\`a del Salento,
                via per Arnesano, 73100 Lecce, Italy}
\address[INFN_LE]{Istituto Nazionale di Fisica Nucleare, Sezione di
                  Lecce, via per Arnesano, 73100 Lecce, Italy}
\address[IHEP]{Key Laboratory of Particle Astrophysics, Institute of High
               Energy Physics, Chinese Academy of Science,
               P.O. Box 918, 100049 Beijing, P.R. China}
\address[INFN_TO]{Istituto Nazionale di Fisica Nucleare,
        Sezione di Torino, via P. Giuria 1, 10125 Torino, Italy}
\address[IFSI]{Istituto di Fisica dello Spazio Interplanetario dell'Istituto
        Nazionale di Astrofisica, corso Fiume 4, 10133 Torino, Italy}
\address[INFN_PV]{Istituto Nazionale di Fisica Nucleare, Sezione di Pavia,
        via Bassi 6, 27100 Pavia, Italy}
\address[HEBEI] {Hebei Normal University, Shijiazhuang 050016, Hebei, China}
\address[INFN_CT]{Istituto Nazionale di Fisica Nucleare, Sezione di Catania,
                 Viale A. Doria 6, 95125 Catania, Italy}
\address[YUNNAN]{Yunnan University, 2 North Cuihu Rd, 650091 Kunming, Yunnan, P.R. China}
\address[DIFTER]{Universit\`a degli Studi di Palermo, Dipartimento di Fisica e Tecnologie
                Relative, Viale delle Scienze, Edificio 18, 90128 Palermo, Italy}
\address[TIBET]{Tibet University, 850000 Lhasa, Xizang, P.R. China}
\address[FISTO]{Dipartimento di Fisica Generale dell'Universit\`a di Torino, via P. Giuria 1, 10125 Torino, Italy}
\address[SHANDONG]{Shandong University, 250100 Jinan, Shandong, P.R. China}
\address[CHENGDU]{South West Jiaotong University, 610031 Chengdu, Sichuan, P.R. China}
\address[PAVIA]{Dipartimento di Fisica Nucleare e Teorica dell'Universit\`a
                 di Pavia, via Bassi 6, 27100 Pavia, Italy}
\address[ING_INNO]{Dipartimento di Ingegneria dell'Innovazione,
                  Universit\`a del Salento, 73100 Lecce, Italy}
\address[IASF_BO]{Istituto di Astrofisica Spaziale e Fisica
        Cosmica di Palermo - Istituto Nazionale di Astrofisica - via Ugo La Malfa 153 - 90146 Palermo -
        Italy}
\address[INFN_VBP]{Istituto Nazionale di Fisica Nucleare - CNAF - viale Berti-Pichat 6/2 - 40127 Bologna - Italy}

\begin{abstract}
The ARGO-YBJ (Astrophysical Radiation Ground-based Observatory at
YangBaJing) experiment is designed for very high energy
$\gamma$-astronomy and cosmic ray researches. Due to the full
coverage of a large area ($5600 m^2$) with resistive plate chambers
at a very high altitude (4300 m a.s.l.), the ARGO-YBJ detector is
used to search for transient phenomena, such as Gamma-ray bursts
(GRBs). Because the ARGO-YBJ detector has a large field of view
($\sim$2 sr) and is operated with a high duty cycle ($>$90\%), it is
well suited for GRB surveying  and can be operated in searches for
high energy GRBs following alarms set by satellite-borne
observations at lower energies. In this paper, the sensitivity of
the ARGO-YBJ detector for GRB detection is estimated. Upper limits
to fluence with 99\% confidence level for 26 GRBs inside the field
of view from June 2006 to January 2009 are set in the two energy
ranges 10$-$100 GeV and 10 GeV$-$1 TeV.
\end{abstract}

\end{frontmatter}

\section{Introduction}
Gamma-ray bursts (GRBs) are very strong gamma-ray photon emissions
from cosmic unpredictable locations in a duration from milliseconds
to tens of minutes. They are the most energetic form of energy
released from a single object in such a short time. The total amount
of light emitted in a GRB is usually a factor of hundreds brighter
than a typical supernova. Using thousands of GRBs detected by
satellite-based detectors, they have been thoroughly investigated in
the keV$-$MeV energy range. They are isotropically distributed in
the sky with a non-thermal origin. According to the time duration,
GRBs are usually classified into long ($>$2 s) and short ($<$2 s)
bursts. Since the first detection by the BeppoSAX satellite for
GRB970228 \cite{Costa}, afterglows are observed after GRBs are
discovered, and this enables the multi-wavelength investigation of
GRBs from the optical band to X-rays. Redshift measurements show
that GRBs occur at cosmological distances (the average redshift of
GRBs observed by the Swift satellite is $z=2.3$ \cite{ref1}). Some
short bursts come from  inside old galaxies with little star
formation, suggesting that they may be originated from mergers of
binary neutron stars or black hole-neutron star systems \cite{ref3}.
Some long bursts are associated with supernovas and confirmed to be
related to deaths of massive stars when central cores collapse to
black holes \cite{ref9}. 

High energy (HE) gamma-ray emissions from GRBs is also observed in
several satellite-born experiments. Energetic Gamma-Ray Experiment
Telescope (EGRET) detected several GRBs with photon energies ranging
from 100 MeV to 18 GeV \cite{Dingu-aip-2001}. Both prompt and
delayed emissions were detected and no high energy cutoff was found
in the spectra. Most importantly, a distinct HE spectral component
was evidently detected in GRB941017 \cite{ref11}. Recently, Fermi
Large Area Telescope (LAT) also detected GeV emissions from
GRB080916C \cite{Tajim-gcn-2008} and 081024B \cite{Nicol-gcn-2008}.
These observations at high energies can place important constraints
on models of emission processes and on parameters of the environment
surrounding the sources of bursts. Very high energy (VHE) emission
up to $\sim$TeV is predicted by several models in both prompt and
afterglow phases \cite{ref12}. Emission at such high energies could
result from electron Self-Synchrotron Compton (SSC) scattering in
either internal or external, forward or reverse shocks. In such
cases, a spectrum with double-peak shape extending into the VHE band
is expected. Some models \cite{ref13} also predict VHE emission due
to decays of secondary $\pi^{0}$ mesons in neutron-rich outflows.
Observations of VHE emission could play a role in discriminating
between these models. The difficulty is that the absorption of the
VHE photons by the Extragalactic Background Light (EBL), due to the
pair production $\gamma+\gamma_{EBL}\rightarrow e^{+}+ e^{-}$,
causes a substantial reduction of the VHE photon flux. This sets a
high upper limit on the sensitivity of a detector used for GRB
search in this energy range. The gamma-ray fluxes from these GRBs
become too small to be detected from current satellite-based
experiments due to their small sensitive areas, so only ground-based
experiments have areas large enough for the detection. 

Search for VHE emission from GRBs  has been done by many
ground-based experiments including extensive air shower arrays and
Cherenkov telescopes. No conclusive detection has been made up to
now, while some positive indications were reported. The Tibet
$AS\gamma$ experiment found an indication of 10 TeV emission in a
stacked analysis of 57 bursts \cite{ref15}. The Milagrito experiment
reported evidence of emission above 650 GeV from GRB970417A with a
chance probability of $1.5\times 10^{-3}$ \cite{ref16}. Evidence of
emission above 20 TeV from GRB920915C was reported about 1 min
earlier than the GRB trigger time at $2.7\sigma$ level by the HEGRA
AIROBICC array, but the position deviated of about $9^{\circ}$
\cite{ref17}. The muon detector GRAND found an excess during
GRB971110 with a chance probability of $3\times 10^{-3}$
\cite{ref18}. Due to limited field of view (FOV), Cherenkov
telescopes, like MAGIC and HESS, can only be operated in follow-up
mode and at least 40 s (usually minutes) are needed to sway the
telescopes to point to the GRB. However, this sets very low upper
limits to the photon fluences at energies around hundreds of GeV
during the afterglow phase \cite{Alber-apj-2007,Aharo-aa-2009}.

With a large FOV ($\sim$2 sr) and high duty cycle ($>$90\%), the
ARGO-YBJ experiment, using a full coverage detector of Resistive
Plate Chambers (RPCs) with area $5600 m^2$, is well suited for GRB
surveying. Following alarms by satellite-borne observations of GRBs,
the ARGO-YBJ detector is used to look for emission from them with a
threshold of a few hundreds of GeV. No significant excess has been
observed yet. In this paper, we place upper limits on the VHE
emission fluences for the GRBs inside the FOV of the ARGO-YBJ
detector. Two models with very different high energy cutoff are
used. The detector performance is investigated using MC simulation.
Based on this, the sensitivity of the ARGO-YBJ detector for GRB
search at different time durations and incident zenith angles is
presented in section 3. The  processes of the data analysis and the
method to search for VHE emission are described in section 4. The
results are reported and discussed in section 5.

\section{The ARGO-YBJ experiment}
The ARGO-YBJ  experiment, a collaboration among  Chinese and Italian
institutions, is designed for VHE $\gamma$-astronomy and cosmic ray
observations and located in Tibet, China at an altitude of $4300m$
a.s.l. The detector consists of a single layer of RPCs, operated in
streamer mode, with a modular configuration. The basic module is a
cluster ($5.7\times7.6 m^{2}$), composed of 12 RPCs
($2.850\times1.225 m^{2}$ each). The RPCs are equipped with pick-up
strips ($6.75\times 61.80 cm^2$) and the fast-OR signal of 8 strips
constitutes the logical pixel (named pad) for triggering and timing
purposes. Hundred and thirty clusters are installed to form a carpet
of $5600 m^{2}$ with an active area of $\sim93\%$. This central
carpet is surrounded by 23 additional clusters (``guard ring") to
improve the core location reconstruction. The total area of the
array is $110\times100m^{2}$. More details about the detector and
the RPC performance can be found elsewhere \cite{ref22}.

The RPC carpet is connected to two independent data acquisition
systems, corresponding to the shower and scaler operation modes.
With the scaler mode, each cluster of the ARGO-YBJ detector counts
the rate of events that have total number of hits $\geqslant 1$,
$\geqslant 2$, $\geqslant 3$ and $\geqslant  4$ every 0.5 s. Using
the count rates, GRBs in the ARGO-YBJ detector FOV (zenith angle
smaller than $45^{\circ}$) are investigated without direction
information \cite{ref23}.

In shower mode, the ARGO-YBJ detector is operated by requiring at
least 20 particles within 420 ns on the entire carpet detector. The
high granularity of the apparatus permits a detailed
spatial-temporal reconstruction of the shower front. The temporal
information is the arrival time of particles measured by Time to
Digital Converters (TDCs) with a resolution of approximately 1.8 ns.
This results in an angular resolution of $0.2^{\circ}$ for showers
with energy above 10 TeV and $2.5^{\circ}$ at approximately 100 GeV
\cite{scia-2007}. In order to calibrate the 18,360 TDC channels, an
off-line method \cite{hehh-2007} is developed using cosmic ray
showers. The calibration precision is 0.4 ns by using 24 h  of data
and the calibration result is  updated every month
\cite{Aiell-app-2009}. 

The central 130 clusters began taking data in June 2006, and the
"guard ring" was merged into the DAQ stream in November 2007. The
trigger rate is $\sim$3600 Hz and the average duty cycle is higher
than $90\%$.

\section{Simulation and sensitivity}
\begin{figure} [ht]
\begin{center}
\mbox{\epsfig{file=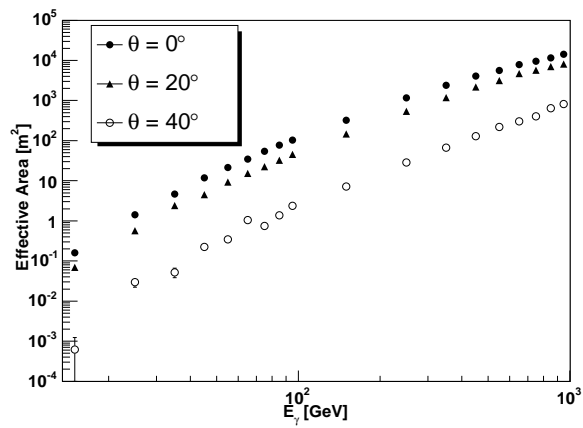,height=6cm}} \caption{ Effective
areas of ARGO-YBJ for gamma rays with $N_{hit}\geq20$ as a function
of the energy for the three zenith angles $\theta=0^{\circ}$,
$20^{\circ}$ and $40^{\circ}$.}
  \label{fig:EffectArea}
\end{center}
\end{figure}

The effective area of the ARGO-YBJ experiment for detecting
gamma-ray showers is estimated by using a full Monte Carlo
simulation driven by CORSIKA 6.502 \cite{Capde-kfk-1992} and
GEANT3-based code ARGO-G \cite{argog}. Five zenith angles
($\theta=0^{\circ},$ $10^{\circ},$  $20^{\circ},$  $30^{\circ}$ and
$40^{\circ}$) are chosen in the simulation and the sampling area is
$300\times300m^2$ around the carpet center. The threshold
multiplicity of hits ($N_{hit}$) which triggers ARGO-YBJ is 20. The
effective areas for this threshold at the three zenith angles
$\theta=0^{\circ},$ $20^{\circ}$ and $40^{\circ}$ are shown in
Fig.$\ref{fig:EffectArea}$.

\vspace{0.5cm}

\begin{figure} [ht]
\begin{center}
\mbox{\epsfig{file=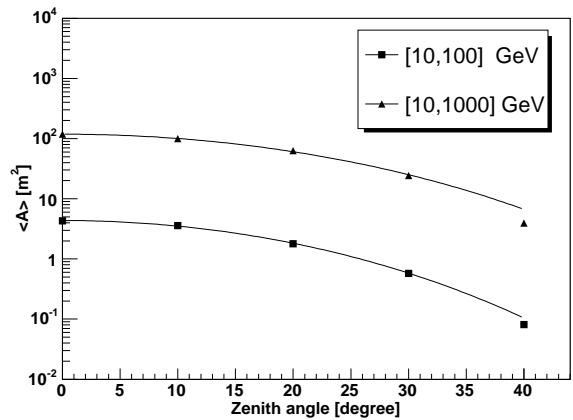,height=6cm}}
 \caption{ The average effective area in two energy ranges after using the event cut listed in Table 1 as a function of
zenith angles.   The lines are the fitting result using the function
$A_{0}\cos^{n}\theta$. } \label{fig:EffectArea2}
\end{center}
\end{figure}

In order to avoid strong absorption of VHE photons by the EBL, two
models of GRB emission with sharp cutoff of their spectra at 100 GeV
and 1 TeV are investigated. Since the multiplicity of hits $N_{hit}$
in an event is related to the shower primary energy, the
optimization of the ARGO-YBJ sensitivity can be done by choosing a
corresponding cutoff on $N_{hit}$. As results of the optimization,
ranges of $N_{hit}$ are found corresponding to the two $E_{cut}$
models as reported in Table 1. Given the ranges of $N_{hit}$, the
corresponding optimal opening angle radii $\phi_{70}$, inside which
70\% of the signal events is included, are reported in Table 1. This
result is almost independent of zenith angle. After using the event
selection listed in Table 1, the average effective areas $\langle A
\rangle$ over the energy ranges from 10 GeV to 100 GeV and from 10
GeV to 1 TeV with a differential spectral index $-$2.0 (this
assumption is used in all the following analyses) are calculated as
functions of the zenith angle $\theta$. They fit a functional form
$\langle A \rangle=A_{0}\cos^{n}\theta$, where $A_{0}=4.36\pm0.04
m^{2}$ and $119.3\pm 0.8 m^2$ are the average effective areas at
$\theta=0^{\circ}$, with the parameter $n=14.26\pm0.11$ and
$10.56\pm0.13$ for the two energy ranges, respectively (see
Fig.$\ref{fig:EffectArea2}$); with these functions the average
effective areas at any zenith may be obtained. At
$\theta=20^{\circ}$, the effective area of the ARGO-YBJ detector is
about $1.8m^{2}$ above 10 GeV if the cutoff energy $E_{cut}$ of the
GRB spectrum is chosen as 100 GeV. On the other hand, it is
$64.3m^{2}$ if $E_{cut}$ = 1000 GeV.

\vspace{0.5cm}

 \vspace{0.5cm}
\begin{figure}[ht]
\begin{center}
 \mbox{\epsfig{file=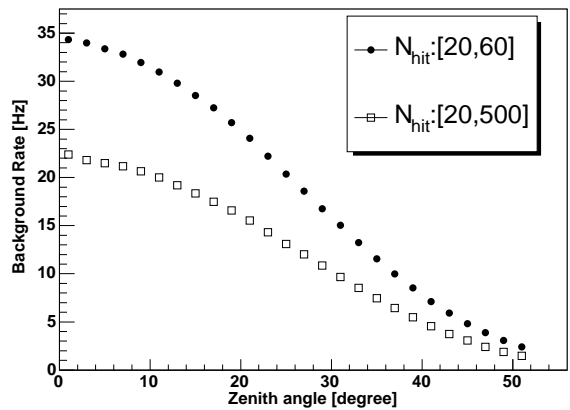,height=6cm}}
 \caption{The background
rate detected in two different data selection conditions listed in
Table 1 as a function of the incident zenith angle. This figure was
made with data taken from one hour before to one hour after the
GRB080903 trigger time, and every point stands for the average value
from different azimuth angles.} \label{fig:BkgRate}
\end{center}
\end{figure}

\begin{table} [ht]
\begin{center}
\caption{$N_{hit}$ ranges and corresponding angular window sizes for
gamma rays.}
\begin{tabular}{lrrr}
\hline
 $E_{cut}$ (GeV) & $N_{hit}$ & $\phi_{70}$ $(^{\circ})$ \\

\hline
100   & 20-60    & $3.8$  \\
1000  & 20-500   & $2.6$ \\
\hline
\end{tabular}
 \label{tab:a}
 \end{center}
\end{table}

\begin{figure}[ht]
\begin{center}
\mbox{\epsfig{file=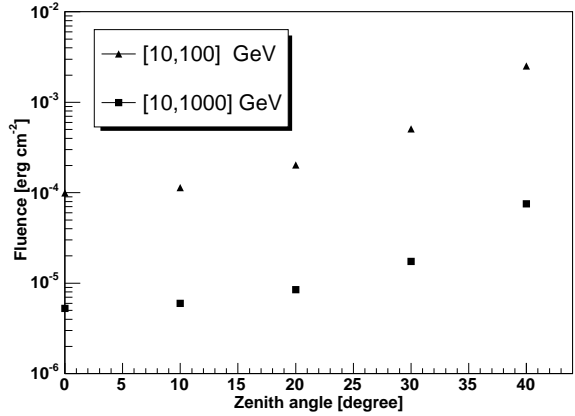,height=6cm}}
 \caption{The $5\sigma$
minimum detectable fluences for GRBs with a duration 12 s in the two
energy ranges [10,100] GeV and [10,1000] GeV for different incident
zenith angles.} \label{fig:GrbSensitive}
\end{center}
\end{figure}

Using measured cosmic ray data, the background event rate (see
Fig.$\ref{fig:BkgRate}$) is estimated according to the selected
$N_{hit}$ ranges reported in Table 1.  Combining this with the
effective areas shown in Fig.$\ref{fig:EffectArea2}$, the minimum
detectable fluences for GRBs requiring a $5\sigma$ excess are
estimated and shown in Fig.$\ref{fig:GrbSensitive}$.  The
sensitivity worsens with the increase of the zenith angle. The GRB
time duration in this figure is fixed at $T=12$ s and the fluence
scales with $T^{1/2}$. We find that the ARGO-YBJ detector has a
sensitivity of $10^{-5}$ $erg/cm^{2}$ in the energy range [10,1000]
GeV for GRBs with a duration of 12 s at $\theta=20^{\circ}$.

\vspace{0.5cm}

\section{Data analysis}
In an angular window surrounding the GRB with the aperture given in
Table 1, the $N_{on}$ events that fall in $N_{hit}$ ranges are taken
as on-source. Among these, the contribution from the cosmic ray
background, $N_{b}$, is estimated by integrating all the events over
2 h around the GRB trigger time, referred as ``direct integral
method" in the literature \cite{Fleys-apj-2004}. The significance of
any excess in the on-source events is calculated using Eq. (17) of
Li and Ma \cite{Litp-apj-1983}. 

As widely discussed, the acceleration mechanism for VHE emission
could be different from that at low energies, so that it could be on
a different time scale. Even if the duration of every burst detected
by satellite is known,  the duration of VHE emission is still
unknown. It is believed that  (a) the most probable emission time is
still in the prompt phase; (b) the emission may be delayed even by
hours, like in the case of the 18 GeV photon in GRB940217, detected
1.5 h after the prompt emission \cite{Hurley}; (c) the high energy
emission could be produced earlier according to some models
\cite{Maxh-hepnp-2003}. The GRB counterpart is first searched in a
window $T_{90}$, which is defined as the time in which 90\% of the
GRB photons is released. If no signal is found, we continue the
search from 1 h  before to 1 h  after the GRB. Time intervals of 1
s, 6 s, 12 s, 24 s, 48 s and 96 s are used in this search with steps
1 s, 2 s, 3 s, 6 s,12 s and 24 s, respectively. If no signal is
found, we set an upper limit to the fluence in $T_{90}$.

Given the observed $N_{on}$ and the expected background $N_{b}$ in
$T_{90}$, the upper limits for the number of events, $N_{UL}$, with
confidence level of 99\% is calculated with Helene's method (Eq.
(10) in \cite{Helen-nim-1983}). Using the effective area and
assuming a differential power law photon spectrum, the upper limit
to the fluence of a GRB in the VHE range is obtained. Guided by the
average spectrum of the four bright bursts observed by EGRET, where
a power law index of $-1.95\pm0.25$ was found over the energy range
from 30 MeV to 10 GeV \cite{Dingu-aip-2001}, an energy spectrum
$dN/dE=CE^{-2}$ is assumed. At first, the normalization constant $C$
is calculated by solving the equation $N_{UL}= \int
A(E)(dN/dE)e^{-\tau_{EBL}}dE$. Then the total fluences can be
obtained by integrating $ \int E(dN/dE)dE$ from 10 GeV to 100 GeV
and to 1 TeV.
 The $\tau_{EBL}$ is the optical depth due to the EBL absorption, and the
optical depths  predicted  by A. Franceschini \cite{Franc-aa-2008}
are used in this work. The absorption factor, which is defined as
the ratio $K(z)= \int A(E)(dN/dE)dE$/$\int
A(E)(dN/dE)e^{-\tau_{EBL}}dE$, for GRBs at redshift $z$=  0.1, 0.5,
1.0 and 2.3 are listed in Table 2. According to these results, the
EBL effect for gamma-rays below 100 GeV is negligible, however, it
could be substantial for farther GRBs in the range [10,1000] GeV.
With the absorption factor $K(z)$ and the average effective area
$\langle A \rangle$ shown in Fig.$\ref{fig:EffectArea2}$, the
parameter $C$ can be directly calculated by solving the equation
$N_{UL}/(K(z)\langle A \rangle)= \int{(dN/dE)}dE$.
\begin{table}
\begin{center}
\caption{The absorption factor $K(z)$ due to the EBL is $K(z)=\int
A(E)(dN/dE)dE$/$\int A(E)(dN/dE)e^{-\tau_{EBL}}dE$.}
\begin{tabular}{lrrrrrrrr}
\hline Energy (GeV) & redshift &$\theta=0^{\circ}$ &
$\theta=10^{\circ}$ &
$\theta=20^{\circ}$ & $\theta=30^{\circ}$ & $\theta=40^{\circ}$ \\
\hline 10-100
&0.5 & 1.08 & 1.08 & 1.08 & 1.08 & 1.09\\
&1.0& 1.39 & 1.39 & 1.39 & 1.39 & 1.40 \\
& 2.3 &3.39 & 3.39 & 3.40 & 3.40 & 3.54 \\
\hline 10-1000
& 0.1 &1.60 & 1.61 & 1.66 & 1.72 & 1.75 \\
& 0.5 &7.92 & 8.02 & 9.27 & 11.13 & 11.66 \\
& 1.0 &25.71 & 26.05 & 31.59 & 39.19 & 41.84 \\
& 2.3 &116.5 & 118.7 & 147.6 & 181.3 & 209.2 \\
\hline
\end{tabular}
 \label{tab:a}
 \end{center}
\end{table}

\section{Results and discussion}
The data used in this work were collected by the ARGO-YBJ experiment
in the periods July 2006$-$July 2007 and November 2007$-$January
2009. Twenty-six GRBs detected by satellites were within the FOV of
the detector array while it was on. Angular resolution, pointing
accuracy and stability of the ARGO-YBJ detector were thoroughly
tested by measuring the Moon shadow in cosmic rays over all
observational periods \cite{Wangb-icrc-2007}. The Crab Nebula and
flares of Mrk421 in years 2006 and 2008 were successfully detected
\cite{Giaco-nima-2008}. 

\begin{figure}[ht]
\begin{center}
\mbox{\epsfig{file=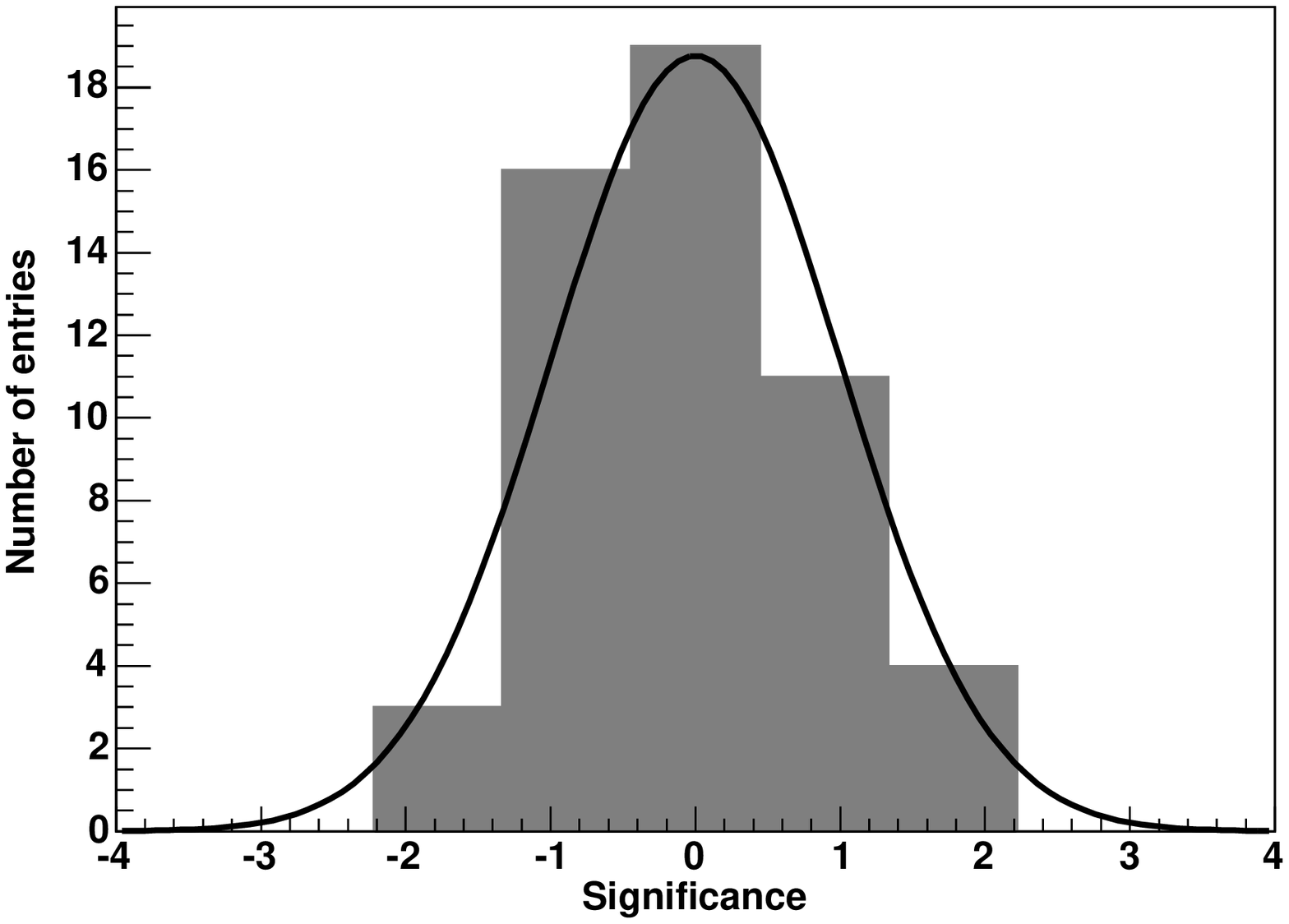,height=4.5cm}}
\mbox{\epsfig{file=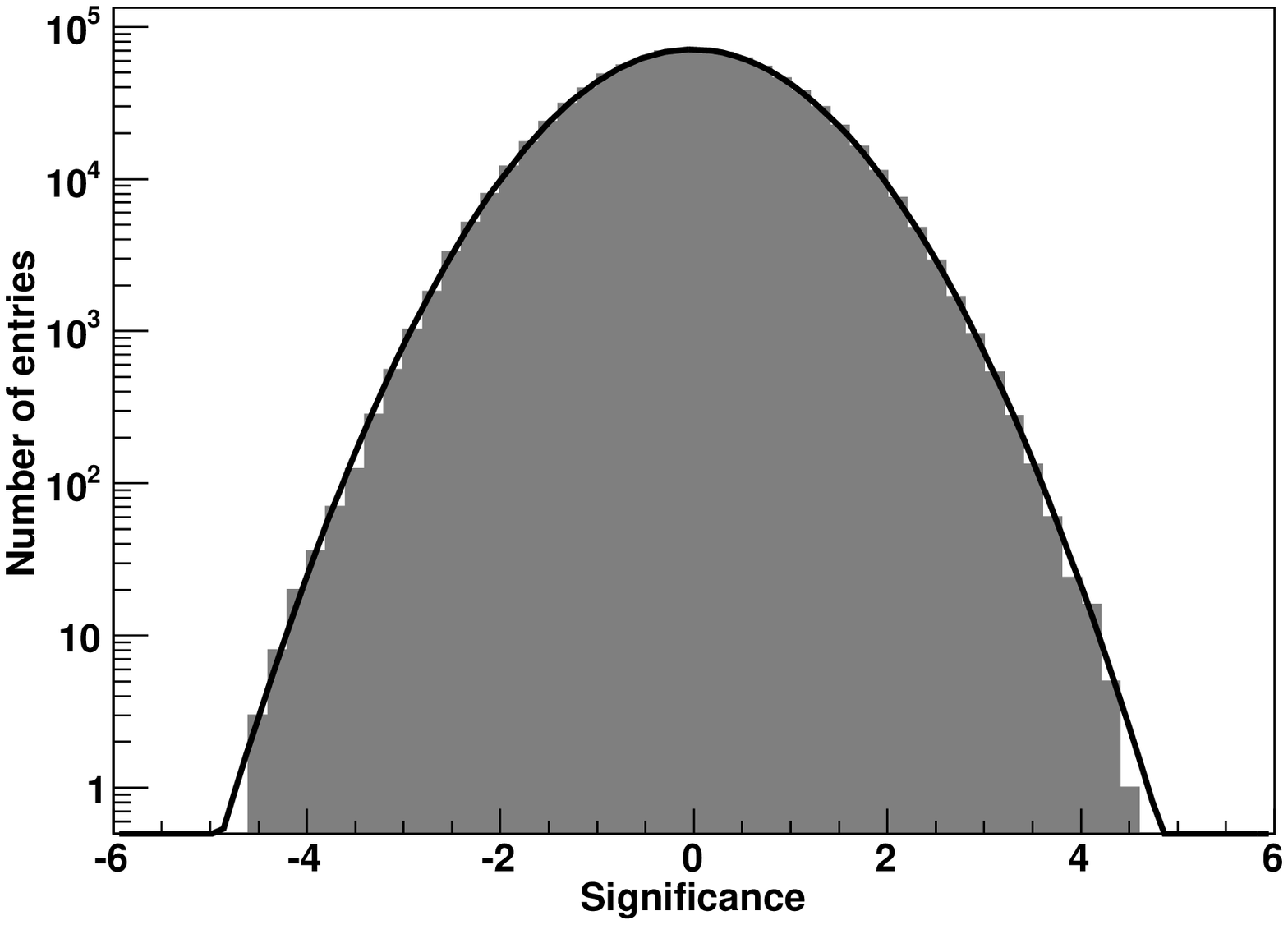,height=4.5cm}}
 \caption{Left: Distribution of the statistical significance of the 26 GRBs
with data available during their prompt phase using both events
selections of Table 1.
Right: Distribution of the statistical significance derived from
two hours of observations around the 26 GRBs analysed with different time
durations. The solid lines are normal Gaussian functions for given
comparison.}
  \label{fig:Significance}
\end{center}
\end{figure}

No significant excess was observed for any of the 26 GRBs, neither
as prompt nor prior/delayed emission (see
Fig.$\ref{fig:Significance}$). Upper limits to fluences in the VHE
range are listed in Table 3. The upper limits for the 6 GRBs with
redshift information have been corrected for EBL absorption, while
the others can be easily corrected using the factors listed in Table
2 assuming $z$=  0.1, 0.5, 1.0 and 2.3. The fluences
in keV bands measured by the satellite experiments are also listed
in Table 3.

\vspace{0.5cm}

\begin{table}
\begin{center}
\caption{List of GRBs in the FOV ($\theta<45^{\circ}$) of ARGO-YBJ and
$99\%$ C.L. fluence upper limits}
\begin{tabular}{lrrrrrrrr}
\hline
 GRB & Satellite & Redshift & $T_{90}$(s) & $\theta$($^{\circ}$) & keV
fluence  & 10$-$100 GeV &
10$-$1000 GeV  \\
 & & & &  &(keV range)&$erg$ $cm^{-2}$&$erg$ $cm^{-2}$&\\

\hline
060714 &Swift  & 2.71   &115  &42.8 &2.9E-6 (15-150) &2.10E-2 &3.11E-2\\
060805B &IPN   & ...   &8    &29.1 &1.1E-4 (30-10000)&1.29E-4 &5.08E-6\\
060807 &Swift  & ...   &43.3 &12.4 &8.5E-7 (15-150) &7.32E-5 &4.23E-6\\
060927 &Swift  & 5.47  &22.6 &31.6 &1.2E-6 (15-150) &6.63E-3 &1.56E-2 \\
061028 &Swift  & ...   &106  &42.5 &9.7E-7 (15-150) &6.23E-3 &1.08E-4\\
061110A &Swift & 0.76  &41   &37.3 &1.1E-6 (15-150) &1.23E-3 &5.32E-4\\
061122 &Integral  & ...   &18   &33.5 &2.3E-5 (20-2000)&4.27E-4 &8.45E-6\\
070306 &Swift  & 1.50  &210  &19.9 &5.5E-6 (15-150) &5.86E-4 &9.63E-4\\
070531 &Swift  & ...   &44   &44.3 &1.1E-6 (15-150) &3.09E-3 &7.82E-5\\
070615 &Integral  & ...   &30   &37.6 &...      &1.42E-3 &3.93E-5\\
071112C &Swift & 0.82  &15   &22.1 &3.0E-6 (15-150) &2.02E-4 &1.04E-4\\
080207 &Swift  & ...   &340&27.7 &6.1E-6 (15-150) &8.95E-4 &3.31E-5\\
080324 &Integral& ...  &13.6   &14.6&...       &8.52E-5 & 5.19E-6\\
080328 &Swift  & ...   &90.6 &37.2 &9.4E-6 (15-150) &1.60E-3 &4.79E-5\\
080602 &Swift  & ...   & 74 & 42.0 &3.2E-6(15-150) &3.31E-3 & 1.32E-4\\
080613B &Swift &...  &105  &39.2 &5.8E-6(15-150) &2.49E-3  &7.32E-5  \\
080726  &AGILE  &...  & 125 &36.7 & ...      &2.38E-3  &5.21E-5 \\
080727C &Swift &...  &79.7 &34.5 &5.3E-6(15-150) &8.15E-4  &3.18E-5  \\
080822B &Swift &...  &64 &40.4 &1.7E-7(15-150) &2.55e-3  &9.46E-5  \\
080903  &Swift &...  &66 &21.5 &1.5E-6(15-150)& 1.73e-4 &7.03e-6\\
081025  &Swift &...  &23 &30.5 &1.9E-6(15-150)&3.75E-4  &1.56E-5\\
081028  &Swift &3.04 &250&29.9 &3.7E-6(15-150)&7.72E-3 &1.25E-2 \\
081105  &Swift &... &10&36.7 &...&4.09E-4 &1.30E-5 \\
081128  & Swift &... &101.7 &31.7 & 2.5E-6(15-150)&6.21E-4&2.05E-5\\
090107  & Swift &... &12.2 &40.1 &2.3E-7(15-150) &7.38E-4&3.21E-5\\
090118  &Swift &...  &16&13.4 &4.0E-7(15-150) &6.68E-5 &3.68E-6\\
 \hline
\end{tabular}
 \label{tab:a}
 \end{center}
\end{table}

\vspace{0.5cm}

\begin{figure}[ht]
\begin{center}
\mbox{\epsfig{file=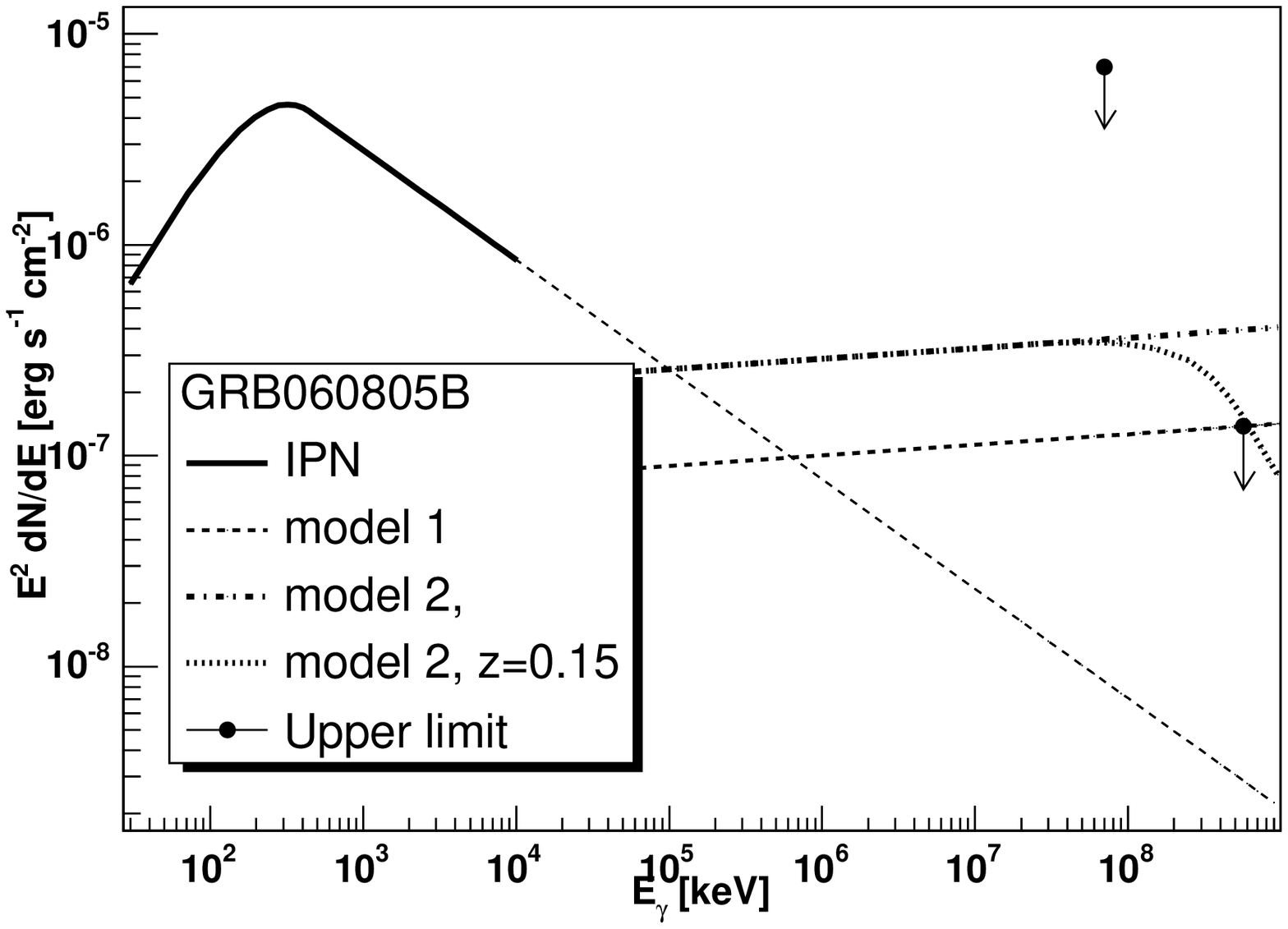,height=6cm}}
 \caption{The spectrum of GRB060805B and the upper
limits obtained by ARGO-YBJ. The solid line is the result of fitting
the IPN data with the Band function and the dotted line is a simple
extrapolation. The upper limits are located at the mean energies of
gamma ray events. For details about model 1 and model 2 see text.}
 \label{fig:GRB060805B}
\end{center}
\end{figure}

It is known that GRB spectra can be well fitted with a Band function
with a break at energy $E_{0}$ mostly between 100 keV and 1 MeV, and
the average index $\alpha$ below the break is $\sim$$-$1 and the
average index $\beta$ above the break is $\sim$$-$2.3
\cite{Preec-apjs-2000}. Most GRBs observed by Swift do not have such
a clear spectral structure with a break since its effective energy
range is often lower than the break energy. GRB060805B was detected
by the Inter Planetary Network (IPN) from 30 keV to 10 MeV and the
result of fitting the data with the Band function ($\alpha$ =
$-$0.66, $\beta$ = $-$2.52 and $E_0$ = 240 keV) is shown in
Fig.$\ref{fig:GRB060805B}$ (solid line). If this GRB spectrum
extends to TeV energies only following the Band function, it will be
compatible with the upper limits obtained by ARGO-YBJ. However, the
spectrum might not extend with a simple power law ($\beta$ =
$-$2.52) above 100 MeV. It might turn to $\beta$ = $-$1.95 like the
four bright bursts detected by EGRET observations
\cite{Dingu-aip-2001}. If EBL absorption is negligible, the limit
found by this work implies that the transition energy should be
above 620 MeV (see model 1 in Fig.$\ref{fig:GRB060805B}$). The other
possibility is that if the transition energy is 100 MeV for this
burst, the source should be farther than z = 0.15 (see model 2 in
Fig.$\ref{fig:GRB060805B}$). In addition, if the SSC mechanism used
by Finke et al. (\cite{Justi-arxiv-2008}) to interpret the spectrum
of GRB 940217 also works on GRB060805B, the limit found by this work
will provide a strong constraint.

In conclusion, we have investigated 26 gamma-ray bursts in the field
of view of ARGO-YBJ in about two years in the GeV$-$TeV energy range
searching for prompt, delayed or prior emission. Since no
significant excess is found, we have set upper limits to fluences
for most of those bursts.

\section{Acknowledgements}
This work is supported in China by the National Natural Science
Foundation of China (NSFC) under the grant 10120130794, the Chinese
Ministry of Science and Technology, the Chinese Academy of Sciences
(CAS), the Key Laboratory of Particle Astrophysics, Institute of
High Energy Physics (IHEP), and in Italy by the Istituto Nazionale
di Fisica Nucleare (INFN). M. Dattoli thanks the National Institute
of Astrophysics of Italy (INAF) for partly supporting her activity
in this work.


\begin{thebibliography}{00}
\bibitem{Costa} E. Costa et al., Nature 387 (1997) 783.
\bibitem{ref1} N. Gehrels, AIP Conf. Proc., Gamma-Ray Bursts 2007 (Santa Fe):3.
\bibitem{ref3} J.S. Bloom   et al., Astrophys. J. 638 (2006) 354.
\bibitem{ref9} A.I. MacFadyen, S.E. Woosley, Astrophys. J. 524 (1999) 262.
\bibitem{Dingu-aip-2001} B.L. Dingus, AIP Conf. Proc. 558 (2001) 383.
\bibitem{ref11} M.M. Gonzalez et al., Nature 424 (2003) 749.
\bibitem{Tajim-gcn-2008} H. Tajima , J. Bregeon , J. Chiang  et al., (2008)  GCN
$\sharp$8246.
\bibitem{Nicol-gcn-2008} N. Omodei  et al., (2008)  GCN
$\sharp$8407.
\bibitem{ref12} P. M\'esz\'aros  et al., Rep. Prog. Phys. 69 (2006) 2259.
\bibitem{ref13} J. Razzaque   et al., Astrophys. J. 650 (2006) 998.
\bibitem{ref15} M. Amenomori   et al., Astron.\& Astrophys. 311 (1996)  919.
\bibitem{ref16} R. Atkins   et al., Astrophys. J. 533 (2000) L119.
\bibitem{ref17} L. Padilla   et al., Astron.\& Astrophys. 337 (1998) 43.
\bibitem{ref18} J. Poirier et al., Phys. Rev. D 67 (2003) 042001.
\bibitem{Alber-apj-2007} J. Albert et al., Astrophys. J. 667 (2007) 358.
\bibitem{Aharo-aa-2009} F. Aharonian  et al., Astron.\& Astrophys. 495 (2009) 505.
\bibitem{ref22} G. Aielli   et al., Nucl. Instr. and Meth. A 562 (2006) 92.
\bibitem{ref23} G. Aielli et al., Astropart. Phys. 30 (2008) 85.
\bibitem{scia-2007} G. Di Sciascio et al., 30th ICRC 4 (2007) 123.
\bibitem{hehh-2007} H.H He et al., Astropart. Phys. 27 (2007) 528.
\bibitem{Aiell-app-2009} G. Aielli et al., Astropart. Phys. 30 (2009) 287.
\bibitem{Capde-kfk-1992} J.N. Capdevielle et al., KfK Report (1992) No.4998.
\bibitem{argog} \url{http://argo.le.infn.it/analysis/argog/}.
\bibitem{Fleys-apj-2004} R. Fleysher   et al., Astrophys. J. 603 (2004) 355.
\bibitem{Litp-apj-1983} T.P. Li, Y.Q. Ma, Astrophys. J. 272 (1983) 317.
\bibitem{Hurley} K. Hurley et al., Nature 372 (1994) 652.
\bibitem{Maxh-hepnp-2003} X.H. Ma et al., HEP \& NP 27 (2003) 973  (in Chinese).
\bibitem{Helen-nim-1983} O. Helene, Nucl. Instr. and Meth. 212 (1983) 319.
\bibitem{Franc-aa-2008} A. Franceschini et al., A \& A 487 (2008)
837.
\bibitem{Wangb-icrc-2007} B. Wang  et al., 30th ICRC 4 (2007) 107.
\bibitem{Giaco-nima-2008} G. D'Al\'{\i} Staiti et al., Nucl. Instr. and Meth. A 588 (2008) 7.
\bibitem{Preec-apjs-2000} R.D. Preece et al., Astrophys. J. S. 126 (2000) 19.
\bibitem{Justi-arxiv-2008} J.D. Finke et al., arXiv:0802.1537v1 [astro-ph].
\end{thebibliography}
\end{document}